\documentclass[reprint,prl,amsmath,amssymb,aps,superscriptaddress,
]{revtex4-1}
\usepackage{graphicx}
\usepackage{dcolumn}
\usepackage{float}
\usepackage{bm}
\usepackage{natbib}
\usepackage{gensymb}

\begin{document}

\preprint{APS/123-QED}

\title{Dipolar thermocapillary motor and swimmer}

\author{V. Frumkin}
\affiliation{Faculty of Mechanical Engineering, Technion - Israel Institute of Technology.}
\author{M. Bercovici}
\email{mberco@technion.ac.il}
\affiliation{Faculty of Mechanical Engineering, Technion - Israel Institute of Technology.}
\affiliation{Department of Mechanical Engineering, The University of Texas at Austin.}

\date{\today}

\begin{abstract}
We study the flow generated by the thermocapillary effect in a liquid film overlaid by a discontinuous solid surface. If the openings in the solid are subjected to a temperature gradient, the resulting thermocapillary flow will lead to a pressure distribution in the film, driving flow in the rest of the system. For an infinite solid surface containing circular openings, we show that the resulting pressure distribution yields dipole flows which can be superposed to create complex flow patterns. For a mobile, finite surface, we show that an inner temperature gradient results in its propulsion, creating a surface swimmer. In addition, we show that these effects can be activated by simple illumination.

\end{abstract}

\maketitle
\textit{Introduction}.-
When an interface between two immiscible fluids is subjected to a non-uniform temperature distribution, the dependence of surface tension on temperature gives rise to tangential stresses which drive fluid motion along the interface. This phenomena, termed the thermocapillary or (more generally) the Marangoni effect, is of central importance in the field of microfluidics due to the dominance of surface forces over body forces at the micro-scale \cite{karbalaei_thermocapillarity_2016}. A non-exhaustive list of applications in this field include the manipulation and production of droplets \cite{sammarco_thermocapillary_1999}, \cite{baroud_thermocapillary_2007}, thermocapillary ratchet flows \cite{stroock_fluidic_2003}, patterning of nanoscale polymer films \cite{dietzel_thermocapillary_2009}, and optical manipulation of microscale fluid flow \cite{garnier_optical_2003}. Theoretical work also suggested enclosing a fluid-liquid interface within micro-devices such as micro-channels \cite{frumkin_creating_2014},\cite{frumkin_liquid_2016} and Hele-Shaw chambers \cite{boos_thermocapillary_1997}, as means of fluid transport.
However, all of the above examples make use of a continuous fluid-liquid interface, while many engineering realizations of such systems may include the interaction of free surfaces with no-slip boundaries. While a number of theoretical studies analyzed flows over superhydrophobic surfaces as a particular case of such interactions, predicting an effective slip \cite{baier_thermocapillary_2010}, \cite{yariv_thermocapillary_2018}, such effectively continuous behavior can only be obtained for periodic systems.
To the best of our knowledge, thermocapillary flows over macro scale, non-periodic solid discontinuities, have not yet been addressed.  

In this work, we study analytically and experimentally the case of a thin liquid film, overlaid by a solid surface containing a circular gap, exposing a limited free surface region. For the case of an infinite and stationary surface, we show that the resulting pressure distribution gives rise to dipole flow in the closed region. Such dipoles can be superposed to create more complex (but well predicted flows). For the case of a finite mobile surface, i.e. a floating annulus, we show that the same phenomenon results in propulsion of the object. Finally, we also show that such propulsion can be actuated by simple illumination, creating photo-activated surface swimmers. 
We here present our theoretical analysis, followed by experimental demonstration and validation.\\

\begin{figure}
\label{fig:Sketch}
\noindent \begin{centering}
\includegraphics[width=20pc]{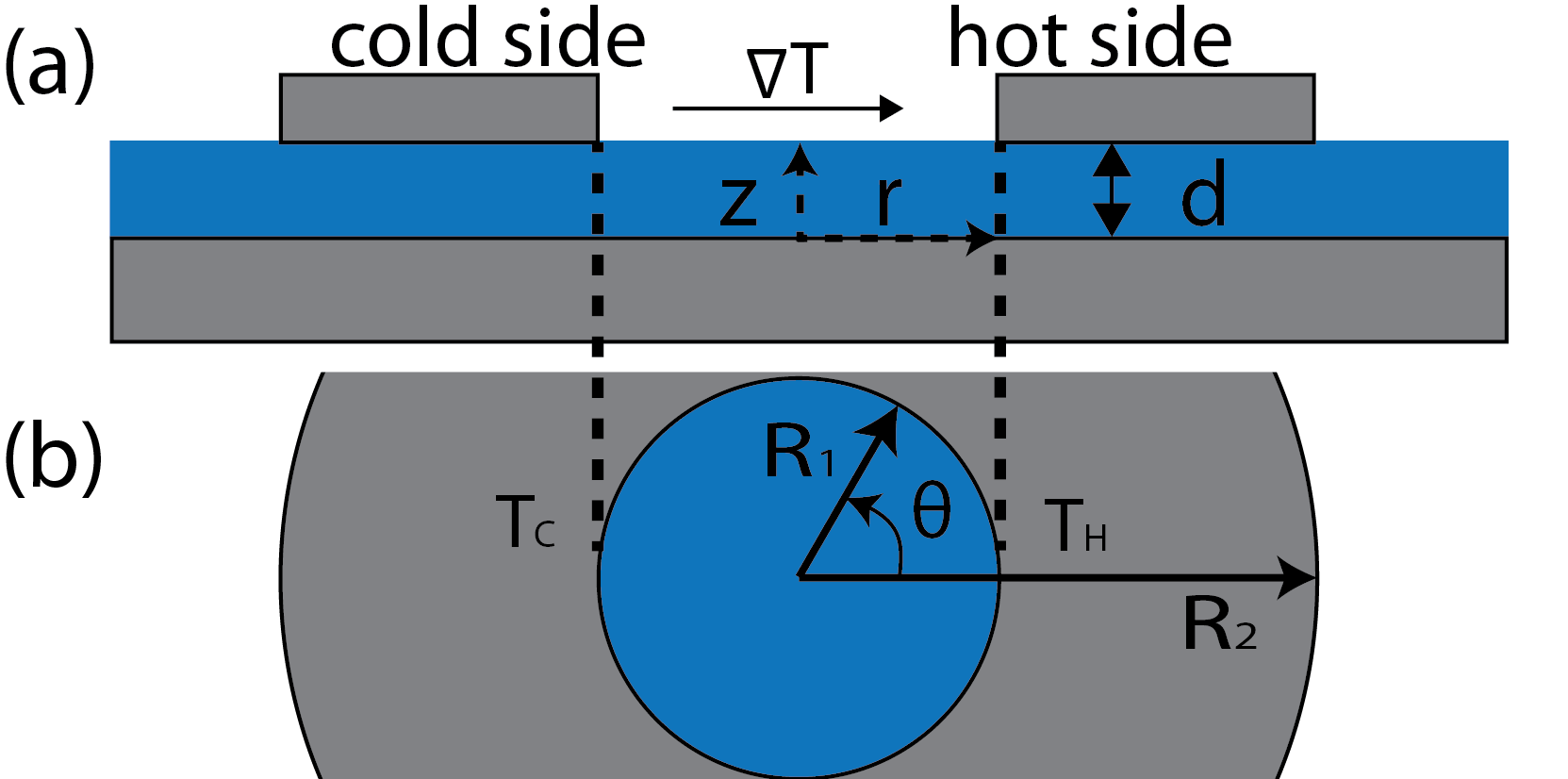}
\par\end{centering}
\caption{Schematic of the system's geometry: (a) Side view of the suspended surface with the circular opening. A temperature gradient is induced from left to right, resulting in a gradient in surface tension in the opposite direction. (b) Top view of the annulus.
}
\end{figure}

\textit{Theoretical analysis}.-
As illustrated in Fig. 1, We consider an annulus with inner radius $R_1$ and outer radius $R_2$, separated by a thin liquid film of height $d$, viscosity $\mu$ and density $\rho$, from a bottom continuous rigid surface. A temperature gradient is induced across the inner opening of the annulus, initiating thermocapillary flow there. We assume the free surface in all regions to be non-deformable, and that the inner radius $R_1$ is much greater than $d$, giving rise to a natural small parameter for the system $\epsilon=d/R_1\ll1$.
We model the system using the steady-state Navier-Stokes and continuity equations in a cylindrical coordinate system, whose origin coincides with the center of the circular opening.

\begin{figure*}[ht]
\label{fig:DDD}
\includegraphics[width=42pc]{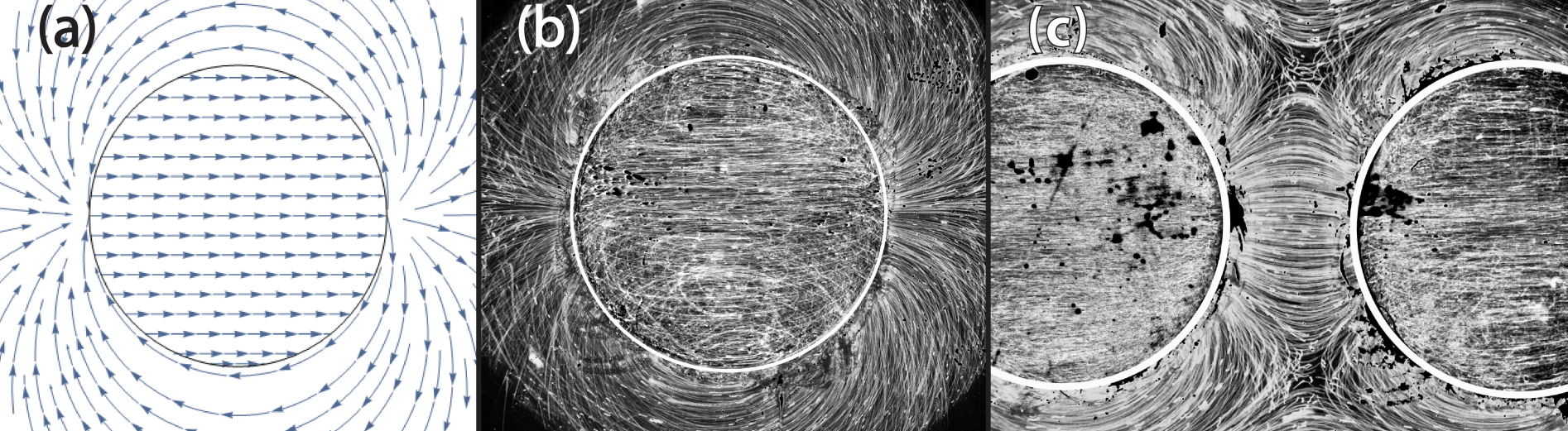}
\caption{
(a) Theoretically predicted streamlines obtained from Eq.(\ref{vel2}) (b) Experimentally measured streamlines for a thermocapillary doublet flow induced in a Hele-Shaw cell through a circular opening of radius $R=4$ mm.
(c) Experimentally measured streamlines for a superposition of two dipole. The flow between the circular openings is directed from left to right, while above and below the openings the flow is directed from right to left. Note the two saddle points of the velocity flied, on the center line between the two openings.}
\end{figure*}

We nondimensionalize radial length by $R_1$, vertical length by $d$, and pressure by $\mu UR_1/d^2$, where $U=\epsilon\sigma_0\triangle/\mu$ is the characteristic thermocapillary velocity, $\sigma_0=\left(\partial\sigma/\partial T\right)$, and $\alpha_{th}$ are the surface tension coefficient, and thermal diffusivity of the liquid, respectively. We nondimensionalize the difference between the temperature in the liquid and the average temperature $(T_H+T_C)/2$ by $\Delta=T_H-T_C$, which is the temperature difference between the hot and the cold sides of the circular region.

We assume a small reduced Reynolds number $\tilde{Re}=\epsilon\frac{\rho Ud}{\mu}\ll1$, which at the leading order in $\epsilon$, yields the linear Stokes and continuity equations 
\begin{gather}
\label{Stokes}
\nabla_{||}p=\frac{\partial^{2}\vec{u}}{\partial z^{2}},\,\,\,\,\,\,\frac{\partial p}{\partial z}=0, \\
\frac{1}{r}\frac{\partial(ru_{r})}{\partial r}+\frac{1}{r}\frac{\partial u_{\theta}}{\partial\theta}+\frac{\partial u_{z}}{\partial z}=0,
\label{Cont}
 \end{gather}
where $\nabla_{||}=\left(\frac{\partial}{\partial r},\,\frac{1}{r}\frac{\partial}{\partial\theta}\right)$ and $\vec{u}=\left(u_{r},\,u_{\theta}\right)$ are respectively, the in-plane gradient and velocity field.
We assume no-slip and no penetration boundary conditions at the lower surface $z=0$, while at the free surface $z=1$, the boundary condition is given by the tangential stress balance,
\begin{equation}
\nabla_{||}T=\nabla_{||}\sigma=\frac{\partial\vec{u}}{\partial z}+\epsilon^2\nabla_{||}\vec{u}\approx\frac{\partial\vec{u}}{\partial z}
\end{equation} 
where $T$ is given by a linear temperature distribution across the inner opening of the annulus
\begin{equation}
    T(\theta)=\frac{r}{2}\cos\theta,\,-\pi\leq\theta\leq\pi,\,\,\,\,0\leq\ r\leq1.
\end{equation}
Integrating Eq. (\ref{Stokes}) in $z$ results in expression for $\vec{u}$, which then can be averaged in $z$ to obtain the mean in-plane velocity, $\left\langle \vec{u}\right\rangle =\int_{0}^{1}\vec{u}dz$, yielding
\begin{equation}
\left\langle \vec{u}\right\rangle =-\frac{1}{3}\nabla_{||}p+\frac{1}{2}H(1-r)\nabla_{||}T,
\label{ug}
\end{equation}
where $H(1-r)$ is the Heaviside step function.
Using the fact that $\nabla_{||}^{2}T=0$ due to the lack of heat sources in our domain, we substitute Eq. (\ref{ug}) into the continuity equation to obtain the Laplace equation for the pressure 
\begin{equation}
\nabla_{||}^{2}p=0.
\end{equation}

The azimuthal symmetry of the problem suggests a solution of the form $p(r,\theta)=R(r)\cos\theta$ which yields a Cauchy-Euler
equation for $R(r)$, with a general solution $R(r)=ar+\frac{b}{r}$, with $a$ and $b$ being constants to be determined. Requiring our solution to be bounded at $r = 0$, yields a solution for the dimensionless velocities and pressure in the three domains, $0\leq r<1;\,\,1<r<\delta;\,\,\delta<r$,  where $\delta=\frac{R_2}{R_1}$
\begin{gather}
\label{vel2}
\left\langle \vec{u}\right\rangle =\begin{cases}
\left(-\frac{a_1}{3}+\frac{1}{4}\right)\left(\cos\theta,\,-\sin\theta\right),\,\,\,0\leq r<1\\
\frac{1}{12}\left((\frac{b_2}{r^2}-a_2)\cos\theta,\,(a_2+\frac{b_2}{r^2})\sin\theta\right),\,\,\,1<r\leq \delta\\
\frac{1}{3}\left((\frac{b_3}{r^2}-a_3)\cos\theta,\,(a_3+\frac{b_3}{r^2})\sin\theta\right),\,\,\, \delta\leq r
\end{cases}\\
p(r,\theta)=\begin{cases}
a_1r\cos\theta,\,\,\,0\leq r<1\\
\left(a_2r+\frac{b_2}{r}\right)\cos\theta,\,\,\,1<r\leq \delta\\
\left(a_3r+\frac{b_3}{r}\right)\cos\theta,\,\,\, \delta\leq r
\end{cases}.
\label{p2}
 \end{gather}
 For a detailed derivation and solution of the governing equations, see supplemental material \cite{noauthor_see_nodate}.

\textit{Thermocapillary dipole}.- For the limiting case when $\delta\rightarrow\infty$, the system corresponds to a Hele-Shaw cell of gap $d$ with a circular opening in its upper plate. We require the velocity to be bounded at infinity, yielding
\begin{gather}
p_{0}(r,\theta)=\begin{cases}
\frac{3}{5}r\cos\theta,\,\,\,0\leq r<1\\
\frac{3}{5}\frac{1}{r}\cos\theta,\,\,\,r>1
\end{cases},
\label{pd}\\
\left\langle\vec{u}_{0}\right\rangle =\begin{cases}
\frac{1}{20}\left(\cos\theta,\,-\sin\theta\right),\,\,\,0\leq r<1\\
\frac{1}{20}\left(\cos\theta,\,\sin\theta\right),\,\,\,r>1
\end{cases}.
\label{vd}
\end{gather}

\begin{figure}[ht]
\label{fig:Track}
\includegraphics[width=20pc]{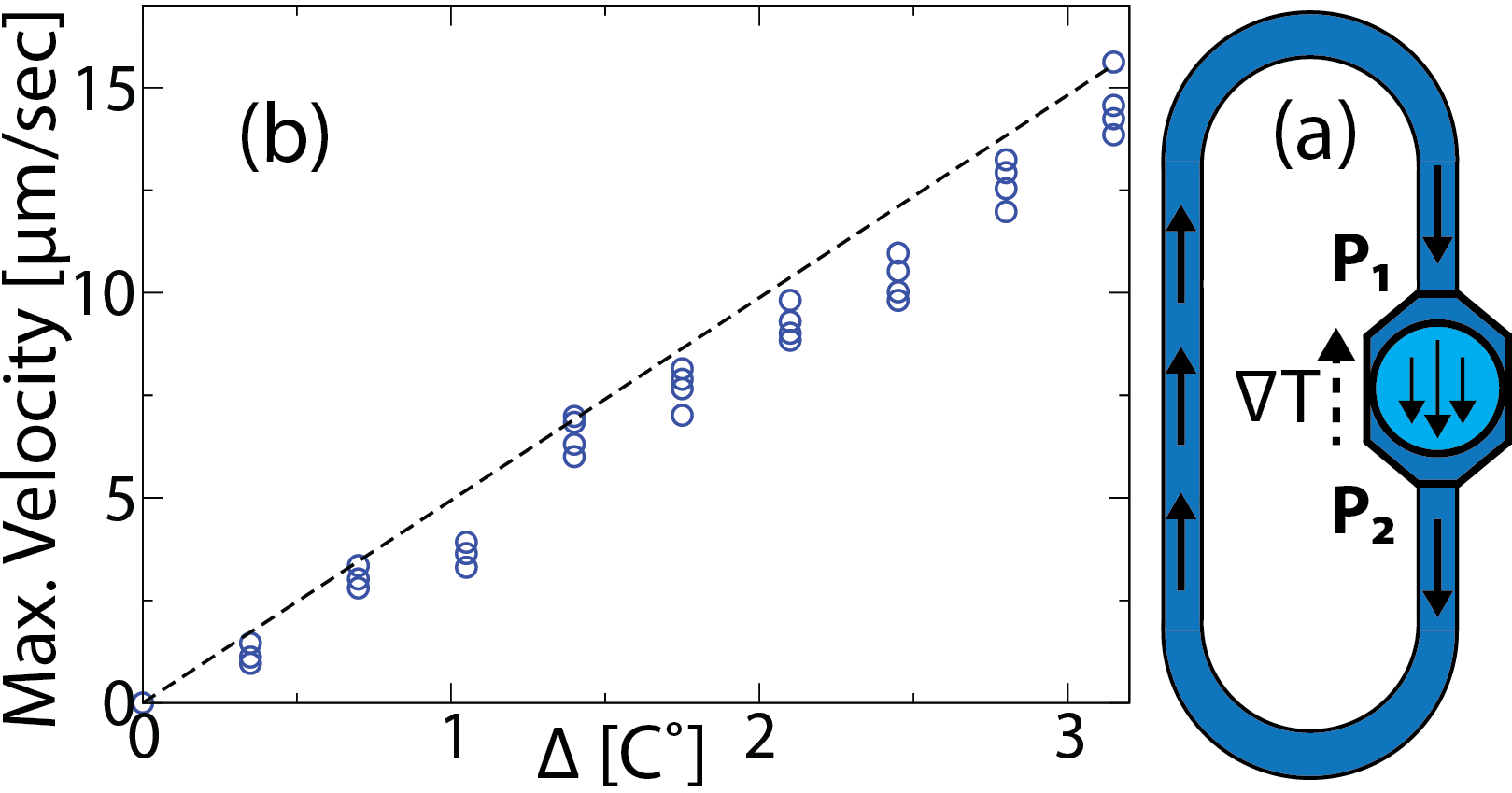}
\caption{(a) Schematic of the TCM experiment - a confined dipole configuration is subjected to a temperature gradient $\nabla T$, inducing thermocapillary flow across the circular opening. The resulting pressure difference drives the liquid through a closed channel.
(b) Experimentally measured maximal velocity as a function of $\triangle$. The dashed line represents the theoretically predicted value given by Eq. (\ref{q}) for $B=0.415$.
}
\end{figure} 

The velocity field described by Eq. (\ref{vd}) corresponds to a doublet flow in a Hele-Shaw cell, driven by thermocapillary stresses on the circular free surface.\\
To implement a thermocapillary dipole, we generated a stable temperature gradient using an $80$ mm aluminum plate, one end of which was heated by a Peltier device while the other end was cooled by a liquid cooler. 
The temperature was controlled using feedback from two thermocouples placed at each end. Such configuration, where the system is heated from below, is equivalent to the solution of imposed temperature at the free surface, under the assumption of a small reduced thermal Peclet $\tilde{Pe}=\epsilon Pe=\epsilon\frac{Ud}{\alpha_{th}}\ll1$, where $\alpha$ is the thermal diffusivity of the liquid. The temperature at the free surface is then modified by a constant coefficient $1/(1+B)$, where  $B=\frac{h_{th}d}{k}$ is the Biot number, $k$ is the thermal conductivity and $h_{th}$ is the heat transfer coefficient.  
On top of the aluminum plate we place a Hele-Shaw chamber consisting of two glass plates separated by distance  $d=0.5$ mm using a PDMS gasket, with one or several circular openings of radius $R=4$ mm in the upper plate (see Fig. 1). The chamber was filled with an aqueous solution of deionized water containing $2\,\mu$m fluorescent beads for flow field visualization.
Fig. 2(a) shows the theoretically predicted streamlines given by Eq. (\ref{vel2}) representing a thermocapillary doublet flow. Fig. 2(b) presents the experimental results for a uniform temperature gradient of $\sim 5$ K/cm, showing good qualitative agreement with theory. A corresponding video is provided in the supplementary material (SV1). Fig. 2(c) depicts a superposition of two thermocapillary dipoles, subjected to the same temperature gradient. The flow between the dipoles is directed from left to right, opposite to the induced temperature gradient, while the return flow above and below the openings is directed along the temperature gradient. These opposing flows give rise to two stagnation points in the flow field where the local velocity vanishes. A corresponding video is provided in the supplementary material (SV2).

Eq. (\ref{pd}) indicates a pressure difference between the two extremes of the circular opening. It is convenient to define a non-dimensional, average pressure difference as:
\begin{equation}
\left<p\right>=\frac{1}{\pi}\int_{-\frac{\pi}{2}}^{\frac{\pi}{2}}\triangle p(1,\theta)d\theta=\frac{12}{5\pi(1+B)},
\end{equation}
where $\triangle p(1,\theta)=p(1,\theta)-p(1,\theta+\pi).$
This suggests that if a dipole unit is confined in the direction perpendicular to the temperature gradient, it can act as a thermocapillary motor (TCM), driving liquids through closed microfluidic circuits. 
To illustrate this, we used a $26$ mm long PDMS channel with a rectangular cross-section of depth $d=0.5$ mm, and width $1$ mm. The channel connected the two ends of an octagonal chamber (see Fig. 3(a)), containing a circular opening of diameter $D=4$ mm, which acted as a TCM unit.
The fluid velocity was measured using PIVlab - Time Resolved Digital Particle Image Velocimetry Tool for MATLAB \cite{thielicke_pivlab_2014}.

For rectangular channels common in microfluidics, the  maximal velocity $u_{m}$ at the center of the channel due to a pressure gradient $\left<p\right>$ is given by the Boussinesq's expression for Poiseuille flow \cite{boussinesq_memoire_1868}
\begin{equation}
\label{q}
u_m=\frac{\left<p\right>}{\mu L}\left(\frac{h^2}{8}-\frac{8}{h}\sum_{n=1}^{\infty}\frac{\sinh(\beta_nl/2)\sin(\beta_nh/2)}{\beta_n^3\sinh(\beta_nl)} \right),
\end{equation}
where $\beta_n=(2n-1)\pi/h$ and $L,h,l$ are the channel's length, height and width, respectively.

Fig. 3b. depicts an experimentally measured maximal velocity in a rectangular channel as a function of the temperature difference $\triangle$ across the circular opening.
The dashed line provides a theoretical comparison with Eq. (\ref{q}) calculated for $B=0.415$. The value of the Biot number was calculated from Eq. (S27) in supplementary material, using the ratio of the measured temperature gradient across the free surface to the imposed gradient on the underlying aluminum plate.

\begin{figure*}[ht]
\label{fig:Swimer}
\includegraphics[width=42pc]{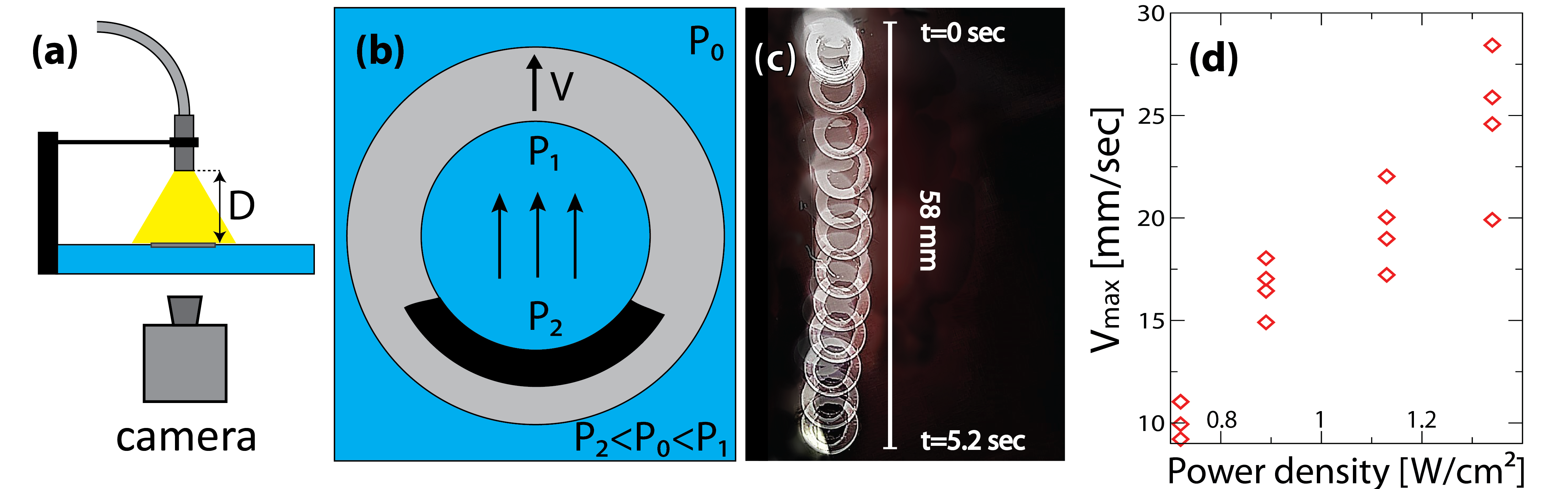}
\caption{(a) Side-view schematic illustration of the surface swimmer experimental setup. The illumination source is located at a distance $D$ from the swimmer, controlling the power density on the swimmer.  (b) Top-view schematic of the surface swimmer. The projected light heats the black stripe on the inner side of the annulus, inducing a temperature gradient across the opening. (c) A set of time-lapse overlaid images showing the swimmer's propagation under uniform illumination. (d) Experimental measurements of the maximal velocity of the swimmer as a function of power density.
}
\end{figure*}

This proof of concept experiment demonstrates that a TCM can serve as a pump, that in contrast to pumping mechanisms such as pressure driven or electro-osmotic flow, is capable of driving flows in closed circuits. The modest flow-rate through the channel can be significantly increased by increasing the cross-section of the channel, shortening its length or increasing the temperature gradient across the TCM. 

\textit{Thermocapillary surface swimmer}.- When delta is finite, the surface can no longer be assumed to be stationary.   In a frame of reference located on the surface, equations (\ref{vel2}) and (\ref{p2}) remain unchanged, except the velocity at infinity does not diminish to zero.  Due to symmetry, we may assume this velocity, $V$, is in the direction of the temperature gradient, i.e. \begin{equation}\lim_{r\rightarrow\infty}\left\langle \vec{u}_{0}\right\rangle  \cdot \frac{\nabla_{||}T}{|\nabla_{||}T|}=-\frac{a_3}{3}=V.
\end{equation}
Solving for the coefficients of Eq. (\ref{vel2}), demanding that the velocity of the free surface, $\vec{u}_{0}(r<1)$, diminishes in the absence of a temperature gradient, $\nabla_{||}T$, yields
\begin{equation}
    V=\frac{ 5 \delta^2-3}{16\delta^2}
\end{equation}
and in dimensional form
\begin{equation}\tilde{V}=V\frac{d\sigma_0\triangle}{R_1\mu}.
\label{us}
\end{equation}
This result, which strictly holds only for case of a shallow liquid film, can be heuristically extended for a swimmer on the surface of an infinite water bath by taking $d$ to be the velocity decay distance, given by \cite{l._napolitano_marangoni_1979}, $d=\sqrt{\frac{R\nu}{U}}$, where $R$ is the length of the free surface and $U$ is the characteristic velocity. For our problem 
\begin{equation}
U= \frac{d\sigma_0\triangle}{R\mu}\,\Longrightarrow d=\left(\frac{R^2\nu\mu}{\sigma_0\triangle}\right)^{\frac{1}{3}}\,\Longrightarrow \epsilon\approx 0.1,
\end{equation}
consistent with the assumptions of the model,
where we have substituted the parameter values used in our experiments, namely $\triangle=4$ and $\delta=\frac{5}{3}$. The resulting predicted velocity of the surface swimmer under these conditions, is  $\tilde{V}\approx 20\, \text{mm/sec}$.

To implement a thermocapillary surface swimmer, we used an annulus made of polypropylene, with thickness $\approx 300\, \mu$m, an inner radius $R_1=3$ mm and an outer radius $R_2=5$ mm. We painted a black stripe on its inner side (see Fig. 4(b)), and used a halogen light source to illuminate the system in order to create a temperature gradient across the inner opening. We recorded the motion of the swimmer by video, and measured its velocity as a function of the projected power density, controlled by varying the distance between the light source and the liquid's surface.

Fig. 4(c) presents an overlaid set of images showing the location of the surface swimmer as a function of time, in response to sudden exposure to a light source concentric with its location at $t=0$. The light absorbed by the black stripe induces a temperature gradient, pushing the surface swimmer in the direction opposite to the location of the stripe. In all experiments we observed an initial acceleration of the swimmer, followed by a decay of its velocity as it leaves the illuminated region. This is shown in supplementary video SV3, whereas supplementary video SV4 shows a case where we follow the swimmer with the light source, resulting in its continuous motion.   
Fig. 4(d) depicts the measured maximal velocity of the surface swimmer, for a given power density of the projected light. We measured velocity of up to $28$ mm/sec, a significant value in the context of micro-flows.  
These results are in good agreement with the estimated value of $U_s\approx20$ mm/sec, obtained from Eq. (\ref{us}), using our experimental parameters and for a typical temperature difference of $\triangle =4\,\text{C\degree}$.
It is interesting to compare the thermocapillary surface swimmer to other methods of Marangoni propulsion, such as the thin rigid circular disk studied theoretically by Lauga and Davis \cite{lauga_viscous_2012}.
There, the disk's propulsion was driven by an external surface tension gradient caused by the release of an insoluble surfactant from part of  the disk's perimeter. In contrast, the thermocapillary swimmer is driven by an internal surface tension gradient, with minimal effect on the surrounding environment. In addition, the thermocapillary
swimmer's velocity is directed oppositely to the temperature gradient, and is aligned with the surface tension gradient, suggesting that a combination of an external and an internal surface tension gradients, should result in an increase of the propulsion velocity.
\\
 
\textit{Summary and discussion}.-
In this paper we demonstrated that thermocapillary flow across a circular opening in a Hele-Shaw type configuration induces a dipole flow inside the Hele-Shaw cell and that such flows could be superposed to obtained 2D flow patterns. A potential extension of this work is the design of flow patterns by distributions of dipoles of different strengths across the flow chambers. This could be achieved using different cavity radii, all subjected to a uniform temperature gradient, or alternatively by localized heating (e.g. by electrodes or illumination) which would allow to control not only the magnitude but also the direction of each dipole. It would also be of interest to explore mass transport in such systems; since the dipole flow has a zero net mass flux, multiple dipoles could be used, for example, to accelerate the mixing between different fluidic chambers in microfluidic applications. We showed that a confined dipole can act as a thermocapillary motor for driving liquids. This mechanism of pumping may be particularly useful in microfluidic applications as it allows to drive flow in a closed circuit.  This in contrast to other standard mechanisms, such as pressure driven flow or electro-osmotic flow, which are inherently directional. The TCM is modular in the sense that it can be positioned in-line in any microfluidic channel. It would thus be of interest to study the effects of various combinations of TCM units (in linear or parallel configurations) on the resulting flow-rate.

Finally, we have demonstrated that a mobile dipole in the form of an annulus can be turned into a thermocapillary surface-swimmer, reaching velocities on the order of $10$ cm/s. While a large body of work exists on a variety of physical mechanisms for swimmers that are suspended in the bulk of the liquid \cite{elgeti_physics_2015}, the unique physics of interfacial phenomena provides new and interesting mechanisms for surface-based propulsion of swimmers. These surface swimmers could potentially be produced on the micro-scale and in large-quantities, and it may be or interest to explore their use as a means for fluidic photo-activation. This would however require further understanding of the viability of such propulsion mechanisms with the decrease in the dimensions of the system.\\

\begin{acknowledgments}\textit{Acknowledgments}.- We thank Dr. Shimon Rubin for fruitful discussions and suggestions. The project was funded by the European Research Council (ERC), grant agreement no. 678734 (MetamorphChip).
\end{acknowledgments}

\bibliography{Dipolar}{}
\bibliographystyle{unsrtnat}

\end{document}